# Full-Dimensional Spatial Light Meta-modulators

Jinwei Zeng, Jinrun Zhang, Yajuan Dong, Jian Wang[*]

Wuhan National Laboratory for Optoelectronics and School of Optical and Electronic Information, Huazhong University of Science and Technology, Wuhan 430074, Hubei, China.

[*] *Correspondence to:* [jwang@hust.edu.cn](jwang@hust.edu.cn)

**Abstract:** The full-dimensional spatial light meta-modulator requires simultaneous, arbitrary and independent manipulation of spatial phase, amplitude, and polarization. It is an essential step towards harnessing complete dimensional resources of light. However, full-dimensional meta-modulation can be challenging due to the need of multiple independent control factors. To address this challenge, here we propose parallel-tasking geometric phase metasurfaces. Indeed, the broadband geometric phase of meta-atoms is divided into several sub-phases, each of which serves as an independent control factor to manipulate light phase, amplitude, and polarization through geometric phase, interference, and orthogonal polarization beam superposition, respectively. Therefore, the macroscopic group of meta-atoms leads to the metasurfaces that can achieve the broadband full-dimensional spatial light meta-modulation. Finally, we fabricate and experimentally demonstrate the meta-modulator that generates special structured light beams with original or modified diffractions, as the signature of spatial phase, amplitude and polarization modulation. This approach paves the way to future wide applications of light manipulation enabled by the full-dimensional spatial light meta-modulators.

A continuous and monochromatic light is an electromagnetic wave defined by its amplitude, phase and polarization[1]. While the conventional optical devices mainly exploit the homogenous amplitude, phase and polarization of light, the recently emerging field of structured light is capable of tailoring the spatial inhomogeneous profile of amplitude, phase and polarization[2-9]. Such "structure" of light can induce the quantized orbital angular momentum (OAM) and vector polarizations to create optical vortices and vector beams, which significantly contribute to the promising applications in super-resolution imaging, optical force manipulating, sensing, quantum information processing, etc[10-17]. Beyond the single-dimensional structuring of light (e.g. spatial amplitude or phase or polarization), the more general structured light accessing multiple dimensions has attracted increasing interest for its newly added degrees of freedom. For example, a high precision holography needs the simultaneous and independent amplitude and phase modulation[18-20], and it can be further segmented and twisted by arbitrary optical vortices to achieve OAM-encrypted holography[21]. Furthermore, the full-dimensional meta-modulation, which we define as the simultaneous, arbitrary and independent control of the spatial amplitude, phase and polarization, is essential to interrogate the comprehensive nature of light. In this scenario, a laudable goal would be to implement the full-dimensional spatial light meta-modulator.

The key step to enable full-dimensional meta-modulation using metasurfaces is to introduce sufficient and effective independent controls of spatial light amplitude, phase and polarization distributions. Previous works have developed several methods to simultaneously control two or more of these light properties. The first method employs the geometric phase (also named as the Pancharatnam-Berry Phase) metasurface, and use both the orientation angle and the structural parameters of the element as control factors[18-20]. The geometric phase element can tune the phase of the orthogonal polarization component of the scattered beam based on the in-plane orientation angle of individual element with broadband property, which can be conveniently described by a Poincare sphere[22-25]. While the orientation angle of the geometric phase elements provides versatile and dispersion-less control over the phase of light, it is only one independent control factor which is insufficient for full-dimensional spatial light meta-modulation. Thus, this method also uses the structural parameters, including size scaling, period/filling fractions etc, which affect the transmission and anisotropy of each element to achieve the amplitude modulation. Therefore, the phase and amplitude modulation can be achieved through varying the in-plane orientation angle and the structural parameters of the element. The other method uses the V, X or Y shape antennas and their structural parameters as control factors[20,26-28]. These antennas can support modes in orthogonal polarizations. Varying the shape, length, width and filling ratio of the antennas can not only decide the overall amplitude (energy) of the scattered light, but also affect the complex coefficient in each orthogonal polarization thus determine the overall polarization and phase of the scattered light. Furthermore, a supercell structure containing several slightly different antennas is designed to introduce interference and realize controllable attenuation for amplitude and phase modulation[18].

For all the aforementioned methods, the control factors always include the delicate structural parameters of the metasurface unitcells. In principle the variation of the structural parameters can provide as many control factors as needed. However, these approaches bring three important limitations in practice. Firstly, the variation of structural parameters may not necessarily enable the complete control over the dimensions of light, i.e. 0~2π phase modulation, 0~1 amplitude modulation, and full Stocks polarization modulation. Secondly, accurate light modulation would require the accurate fabrication of the metasurfaces with extreme fidelity. The nano-fabrication may be very difficult for the structure with acute angle and sharp turn. Especially, when the relationship between the structural parameters and the amplitude, phase and polarization modulation is inexplicit, a small deviation of fabrication may cause a significant modulation error. Lastly, the light modulation controlled by structural parameters is usually very sensitive to wavelength and susceptible to chromatic dispersion. Therefore, the approach of varying structural parameters can only accurately modulate the light in a single wavelength with a very narrow bandwidth.

To overcome these limitations, here we propose a full-dimensional meta-modulation method based on the parallel-tasking geometric phase metasurface. The parallel-tasking means every metasurface unitcell aims to several different tasks simultaneously rather than a single task, which eventually provides sufficient control factors to modulate multiple dimensions of light. In particular, we use the pure geometric phase metasurfaces and only manipulate the orientation angles of the element without varying structural parameters to preserve its broadband property. Importantly, we define sub-angles from the total orientation angles of each element, which control the sub-geometric phases that eventually lead to independent phase, amplitude and polarization modulation, through geometric phase, interference, and orthogonal polarization light superposition principles, respectively. Therefore, the metasurfaces, as the collective group of such parallel-tasking geometric phase elements, are able to achieve the full-dimensional spatial light meta-modulator.

In this work, we design the plasmonic nanobar as the geometric phase element, in which the orientation angle can be divided into three sub-angles to control three sub-phases shown in Fig. 1. We derive an explicit relation between the sub-phases and the modulated phase, amplitude polarization of the transmitted light field. Based on this principle we design and fabricate the metasurface samples to generate special structured light beams with original or modified diffraction as the important proof of principle experimental demonstration of the full-dimensional spatial light meta-modulator.

## Results

## Parallel-tasking principle

We elaborate the parallel-tasking principle for the full-dimensional meta-modulator as shown in Fig. 1. The goal is to generate an arbitrary transmitted beam (e.g. arbitrary amplitude, phase and polarization) at the near field of the metasurface transmission plane from an incident linear polarization (LP) plane wave by the parallel-tasking geometric phase metasurface (Fig. 1a). We set a LP incident plane wave normally transmitting from the substrate side to the metasurface (Fig. 1a). For mathematic convenience, we use circular polarization (CP) basis to express the electric fields of the incident beam and the produced arbitrary beam as follows

$$\mathbf{E}_{\mathbf{in}}(x,y) = A_0 \begin{bmatrix} \sqrt{2}/2 \\ \sqrt{2}/2 \end{bmatrix}, \quad \mathbf{E}(x,y) = A(x,y)e^{i\varphi(x,y)} \begin{bmatrix} |E_l(x,y)|e^{-i\delta(x,y)/2} \\ |E_r(x,y)|e^{i\delta(x,y)/2} \end{bmatrix}, \quad (1)$$

where $A_0$ is the amplitude of the incident LP plane wave, the vector represents the CP basis, $A(x,y)$, $\varphi(x,y)$, $E_l(x,y)$, $E_r(x,y)$, $\delta(x,y)$ are the amplitude profile, phase profile, unit amplitude profile for the left-hand circular polarization (LCP) component, unit amplitude profile for the right-hand circular polarization (RCP) component, and relative phase difference profile between LCP and RCP components of the produced beam, respectively.

We remind and emphasize the geometric phase principle as an essential physics of the proposed metasurface. Any periodic anisotropic structure can serve as a geometric phase element[25]. Under the CP basis, the modulated transmission phase $\varphi$ of the orthogonal polarization component (with regards to the incident polarization) from a geometric phase element is twice of the element's in-plane rotation angle $\theta$ as $\varphi = \pm 2\theta$. The positive and negative sign of the modulated phase is determined by the RCP and LCP handedness of the transmitted orthogonal polarization component, respectively[25]. Here, we use a rectangular nanobar in square lattice to illustrate the geometric phase element. Therefore, the design of the geometric phase metasurface is to determine the orientation angle of each nanobar in the metasurface. However, as explained previously, the geometric phase itself only directly modulates the phase of light, which alone does not explicitly control the full dimensions of light.

To address this problem, the central concept of this work, i.e. the parallel-tasking, is to enable simultaneous and independent tasks of one geometric phase element to achieve sufficient control factors for full-dimensional meta-modulation. For this purpose, as illustrated in Fig. 1b, we introduce three sub-angles from the total orientation angle ($\theta$) of a geometric phase element, i.e. $\theta_\varphi$, $\theta_i$, $\theta_s$, for the parallel tasks of direct phase manipulation, amplitude modulation by interference, and polarization modulation by orthogonal polarization beam superposition, respectively. Based on such mechanism, we need to determine three sub-angles and thus the total orientation angle of each geometric phase element.

First, for phase modulation we introduce the sub-angle $\theta_\varphi$ and use the geometric phase principle, as shown in Fig. 1c. As explained previously, it is straightforward to show that $\varphi = \pm 2\theta_\varphi$ while the sign is dependent on the handedness of the transmitted orthogonal CP component.

Second, for amplitude modulation we introduce the sub-angle $\theta_i$ and use the interference mechanism. As shown in Fig. 1d, two adjacent elements with different sub-angles $\theta_{1i}$ and $\theta_{2i}$ introduce the interference phases $\varphi_{1i} = 2\theta_{1i}$ and $\varphi_{2i} = 2\theta_{2i}$, respectively, to the transmitted RCP components. Remarkably, due to the subwavelength period of the metasurface, we consider the transmitted beams from two adjacent elements overlap each other. Therefore, the final interference beam has the expression of $\cos[(\varphi_{1i} - \varphi_{2i})/2]e^{i[(\varphi_{1i}+\varphi_{2i})/2]}$. Here, we let $\theta_{1i} = -\theta_{2i}$, and it means the interference sub-angles enable the independent amplitude modulation.

Third, for polarization modulation we introduce the sub-angle $\theta_s$ and use the superposition of two orthogonally polarized beams[29,30]. As shown in Fig. 1e, the sub-angle $\theta_s$ provides a phase gradient to the horizontally adjacent geometric-phase elements to diffract the transmitted beam. We set the sub-angles $\theta_s$ in the odd and even rows of the metasurface to be exactly opposite. Therefore, under LP illumination which can be decomposed as the superposition of LCP and RCP beams, the odd and even rows of the metasurface will diffract the RCP and LCP components of the transmitted beam at the same diffraction angle, respectively. Again, due to the subwavelength period of the metasurface, we consider the diffracted LCP and RCP beams from the odd and even rows as one overlapped beam. Since the interference sub-angles in the odd and even rows can provide independent amplitude and phase modulation of the two orthogonally polarized transmitted beams, they can achieve independent and complete polarization modulation.

Based on the above parallel-tasking mechanisms (geometric phase, interference, orthogonal superposition) enabling independent and simultaneous phase, amplitude and polarization modulation, the full-dimensional spatial light meta-modulator can be achieved.

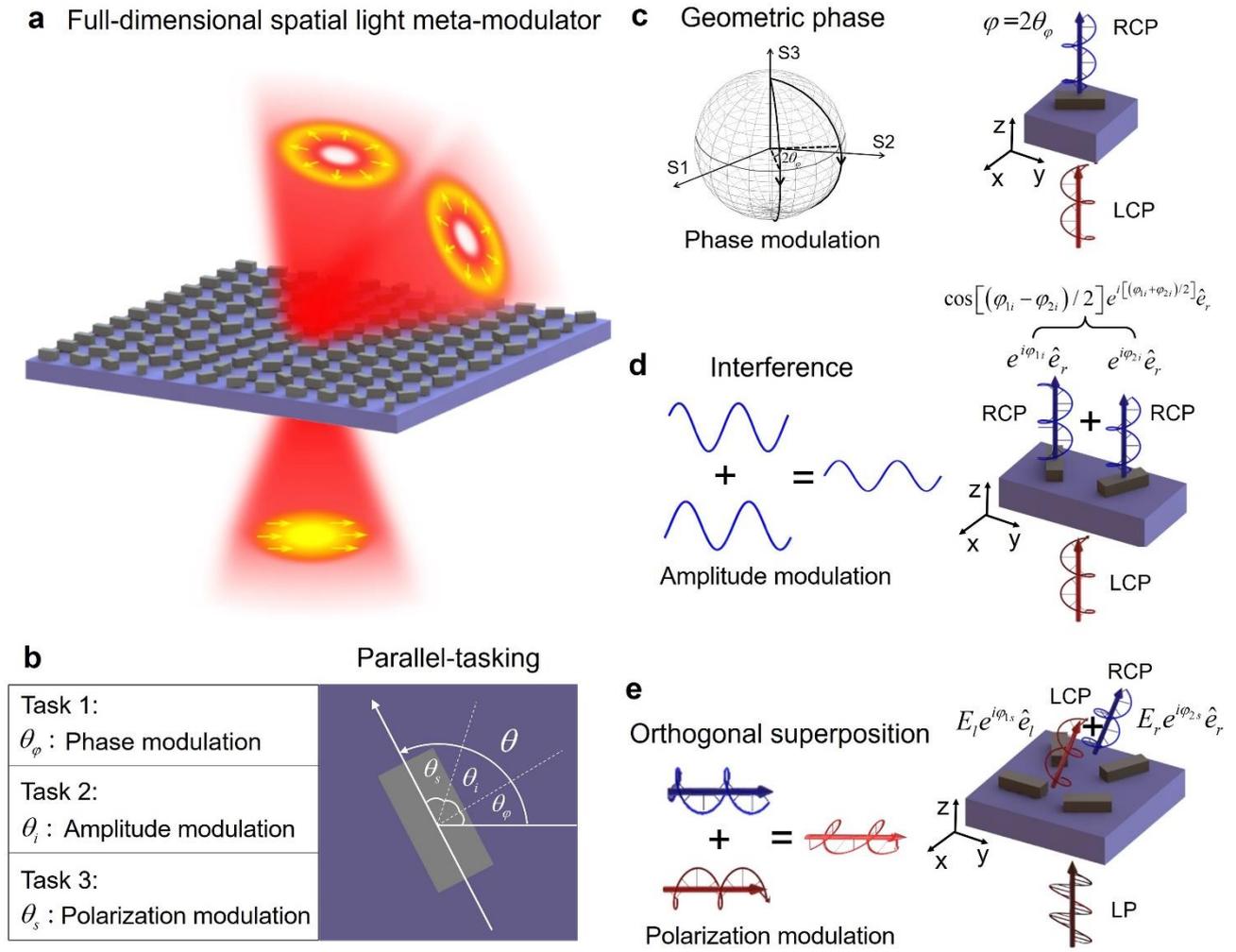

**Figure 1. Concept and principle of full-dimensional spatial light meta-modulators. a** The conceptual schematic of the full-dimensional spatial light meta-modulator. The incident linear polarization (LP) plane wave is converted to a full-dimensional spatial light meta-modulation beam (e.g. selective multi-level diffraction of the vector beam) by the parallel-tasking geometric phase metasurface. **b** The schematic of the parallel-tasking mechanism of a single geometric phase element. The total orientation angle of the element is divided into three sub-angles for parallel tasks ($\theta_\varphi$ for direct phase manipulation, $\theta_i$ for amplitude modulation by interference, $\theta_s$ for polarization modulation by orthogonal polarization beam superposition). **c** Task 1: geometric phase for phase modulation. **d** Task 2: interference for amplitude modulation. **e** Task 3: orthogonal superposition for polarization modulation.

## Plasmonic nanobars forming parallel-tasking metasurfaces

Based on the supercell principle we design the geometric phase metasurfaces to demonstrate the full-dimensional spatial light meta-modulator as promised. It contains two parts: the unitcell structure as the fundamental building block of the metasurfaces, and the unitcell arrangement which enables the parallel-tasks of the full-dimensional spatial light meta-modulator.

The first step is to determine the particular nano-structure as the unitcell. Since the proposed parallel-tasking mechanism only dictates the orientation of each geometric phase element in the supercell, it does not restrict the shapes, sizes and materials of the element. Therefore, any anisotropic nano-structure that follows the geometric phase principle can serve as an eligible geometric phase unitcell in this work. Here, as shown in Fig. 2a, we design the plasmonic nanobar antenna as the unitcell of the geometric phase metasurface to construct the parallel-tasking metasurfaces. Although the dielectric elements reported in the previous works excel in better polarization conversion efficiency (PCE)[4,31,32], the plasmonic element is more convenient to fabricate and in this work we choose it as an excellent candidate for the proof of principle demonstration. We design the gold nanobar with its thickness, width and length as 100, 100 and 200 nm, respectively. These gold nanobars are arranged in a square lattice with the period of 300 nm. Fig. 2b shows the measured scanning electron microscopy (SEM) image of a fabricated metasurface sample in this work. Shown in Fig. 2c is the zoom-in view of the measured SEM image. In consistency of the supercell design, we define the plane of the metasurface as the x-y plane, and the incident light transmits through the metasurfaces from the sample side along the z-axis. Also, we define the transverse-electric (TE) or transverse-magnetic (TM) polarizations as the incident electric field **E** or magnetic field **H** is parallel to the long side of the gold antenna, respectively. We use a finite difference time domain (FDTD) method to calculate the electromagnetic response of the gold nanobar antenna unitcell from 400 to 760 nm (covering the full visible light range). For this purpose, we simulate the transmission of such unitcell under TE and TM linear polarizations, as shown in Fig. 2(d). It indicates an obvious anisotropy between the two axes of the gold nanobar antenna. Then, we calculate the PCE as shown in Fig. 2e, which is defined as the energy ratio of the converted polarization component to the incidence light. Lastly, we simulate the transmission phase of one unitcell under CP basis, and plot the phase shift in the orthogonal CP component in the transmitted beam versus the rotating angle of the nanobar antenna at 633 nm, as shown in Fig. 2f. It clearly shows that the phase shift is about twice as the rotating angle of the element, which is in excellent agreement with the geometric phase principle. These results confirm the designed gold nanobar as an eligible element to construct the geometric phase metasurfaces.

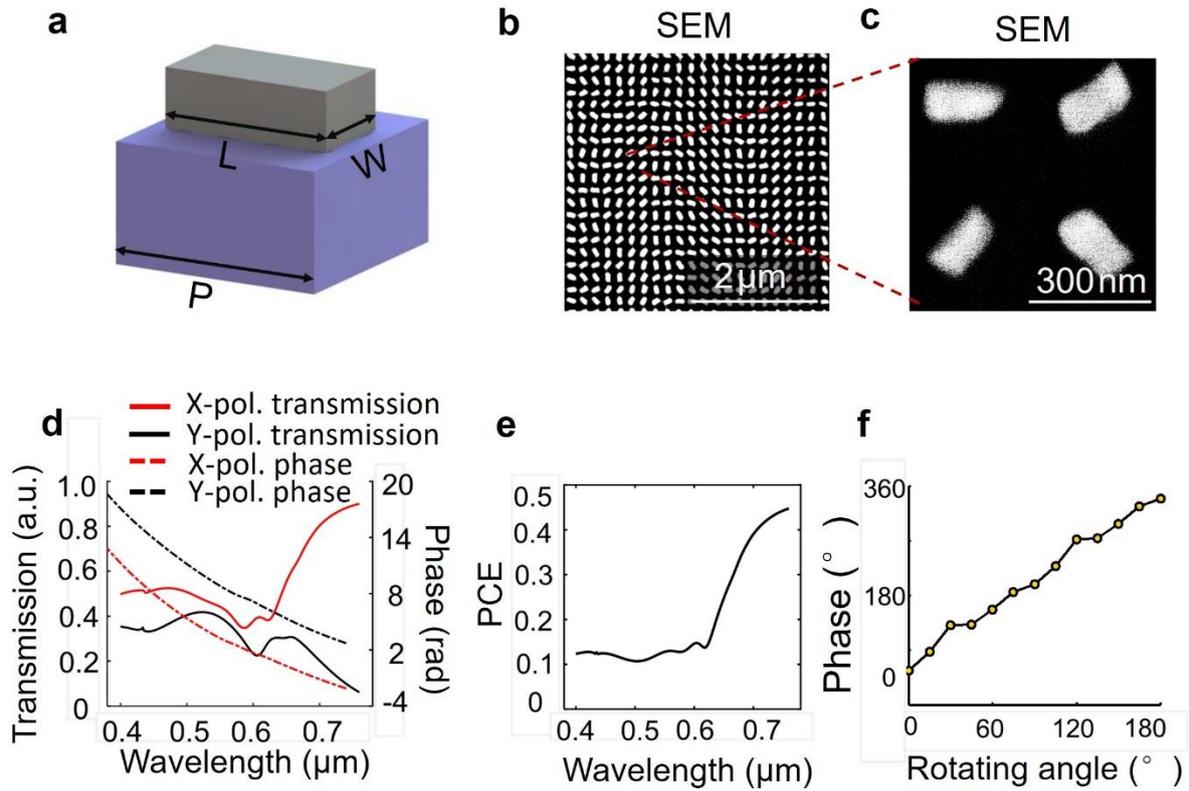

**Figure 2. Plasmonic nanobars forming parallel-tasking metasurfaces and their electromagnetic response. a** The schematic diagram of the plasmonic nanobar antenna unitcell. **b** Measured scanning electron microscopy (SEM) image of the fabricated metasurface sample. **c** Zoom-in view of the fabricated metasurface. **d** Simulated transmission and phase responses of the plasmonic nanobar under TE (X-pol) and TM (Y-pol) linear polarizations from 400 to 760 nm. **e** The polarization conversion efficiency (PCE) of the plasmonic geometric phase element from 400 to 760 nm. **f** Phase shift versus rotating angle of the plasmonic nanobar element.

The second step is to determine the rotating angle of each element of the parallel-tasking metasurfaces. Here, we illustrate the generation of optical vortices, vector beams, vector vortices, and the vector beams in selective multi-level diffraction as several typical examples, giving the experimental demonstration of the full-dimensional spatial light meta-modulator.

## Spatial phase modulation for generating optical vortices

We first design and fabricate the metasurface to generate optical vortices. The optical vortex is a structured light with helical phase front, which has a phase singularity at the beam center[33]. The degree of the phase twisting can be described by the quantized topological charge of the orbital angular momentum (OAM)[33]. It needs a spiral phase profile modulation to generate the optical vortex. As shown in Figs. 3a and 3b, the OAM-carrying optical vortex beam presents a doughnut-shape intensity profile and a spiral interferogram formed by the interference with a reference Gaussian beam.

**Spatial polarization modulation for generating vector beams**

We also design and fabricate the metasurface to generate vector beams. The typical vector beams include azimuthally polarized beam (APB) and radially polarized beam (RPB), which have a polarization singularity at the beam center[12]. The fabricated metasurface enables the corresponding polarization profile modulation to generate the APB and RPB. As shown in Figs. 3c-3f, the vector beams also exhibit a doughnut-shape intensity profile, and they can be further characterized and verified with a polarizer. When rotating the polarizer, the double-lobe shape intensity distribution after the polarizer rotates accordingly[12].

**Spatial phase and polarization modulation for generating vector vortices**

We then fabricate the metasurface to generate vector vortices. The vector vortex can be regarded as the combination of the optical vortex and vector beam, which simultaneously has helical phase front and vector polarization[34]. It accesses both the spatial phase and spatial polarization dimensions. Note that for the vector vortex with the OAM charge of 1, its phase and polarization singularities are counteracted at its beam center, thus its beam profile shows a bright center. The vector vortices can be characterized and verified using polarizer, together with LCP and RCP interferograms, as shown in Figs. 4a-4d.

**Full-dimensional spatial light meta-modulation (amplitude, phase, polarization)**

We finally fabricate the metasurface to generate a special light beam with full-dimensional spatial light meta-modulation. We choose a typical example of the selective multi-level diffraction of the vector beam. The selective multi-level diffraction means to modulate the amplitude of the selective diffraction orders, which requires a simultaneous amplitude and phase modulation[20,27]. Therefore, the selective multi-level diffraction of a vector beam requires a simultaneous amplitude, phase and polarization modulation, thus it provides a direct demonstration of the full-dimensional spatial light meta-modulator. In this work, we produce the APB and RPB in the 1st and 2nd diffraction order and make them equally bright, as clearly indicated in Figs. 5a and 5b.

We plot the experimental and simulation results in Figs. 3-5 for the generation of optical vortices, vector beams, vector vortices and vector beams in selective multi-level diffraction. One can clearly see that the experimental results are in excellent agreement with the theoretical expectations and simulation results, which provides a solid demonstration for the proposed parallel-tasking full-dimensional spatial light meta-modulator.

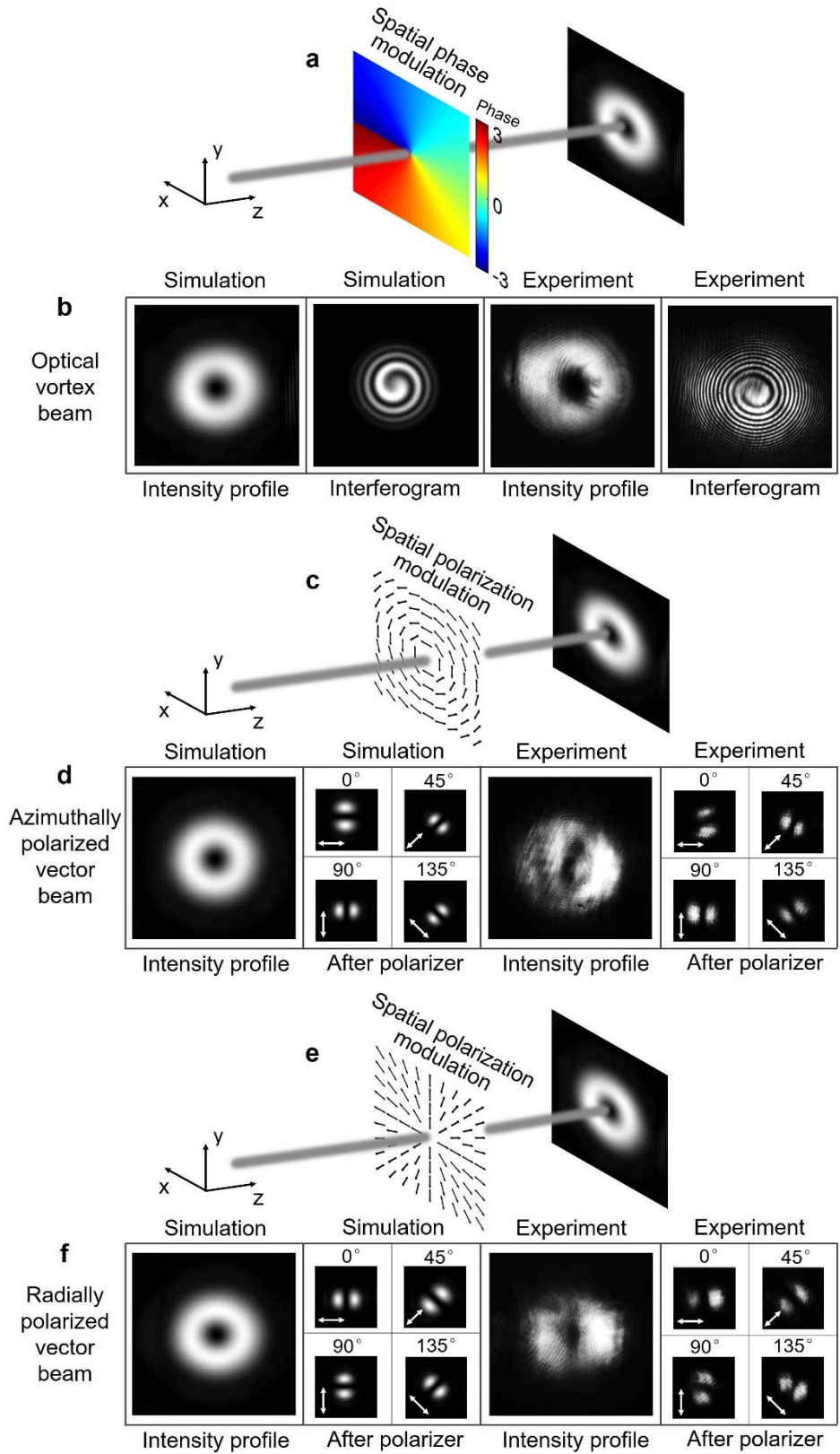

**Figure 3. Experimental and simulation results for generating optical vortices and vector beams with the designed and fabricated metasurfaces. a**, **b** Illustration (**a**) and results (**b**) for generating the optical vortex beam (spatial phase dimension). **c**, **d** Illustration (**c**) and results (**d**) for generating the azimuthally polarized vector beam (spatial polarization dimension). **e**, **f** Illustration (**e**) and results (**f**) for generating the radially polarized vector beam (spatial polarization dimension).

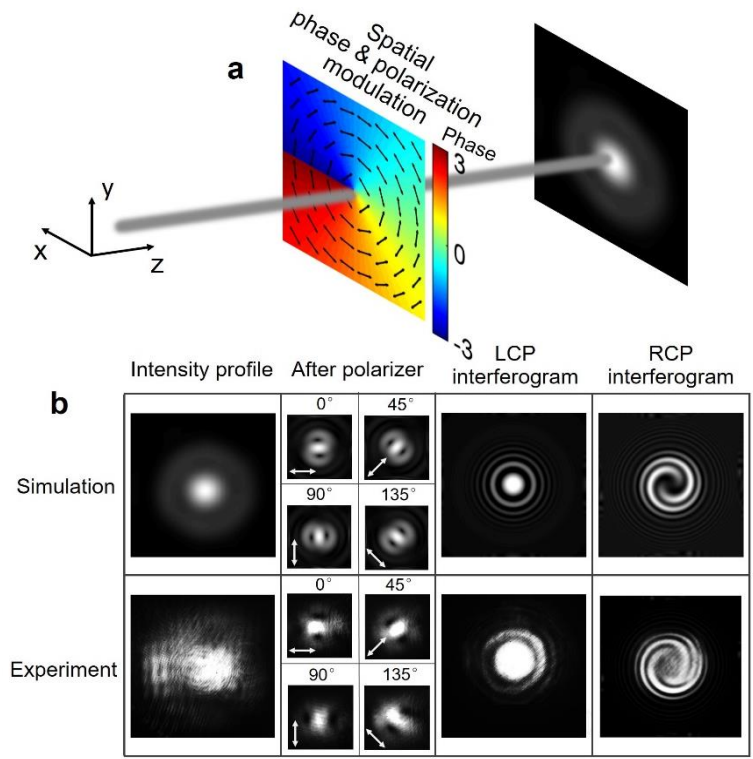

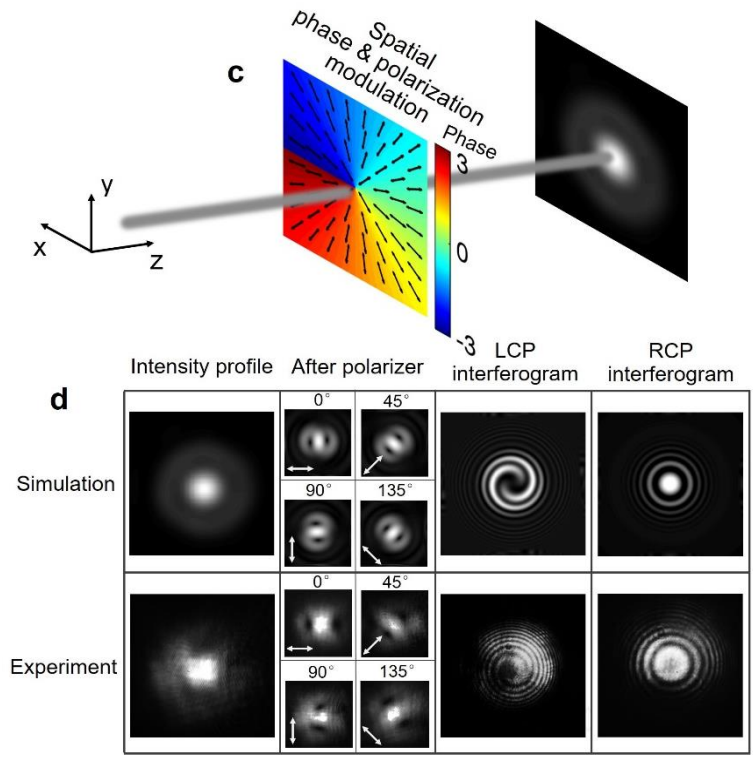

**Figure 4. Experimental and simulation results for generating vector vortices with the designed and fabricated metasurfaces. a, b** Illustration (**a**) and results (**b**) for generating vector vortex with azimuthal polarization distribution and helical phase distribution (spatial phase and polarization dimensions). **c, d** Illustration (**c**) and results (**d**) for generating vector vortex with radial polarization distribution and helical phase distribution (spatial phase and polarization dimensions).

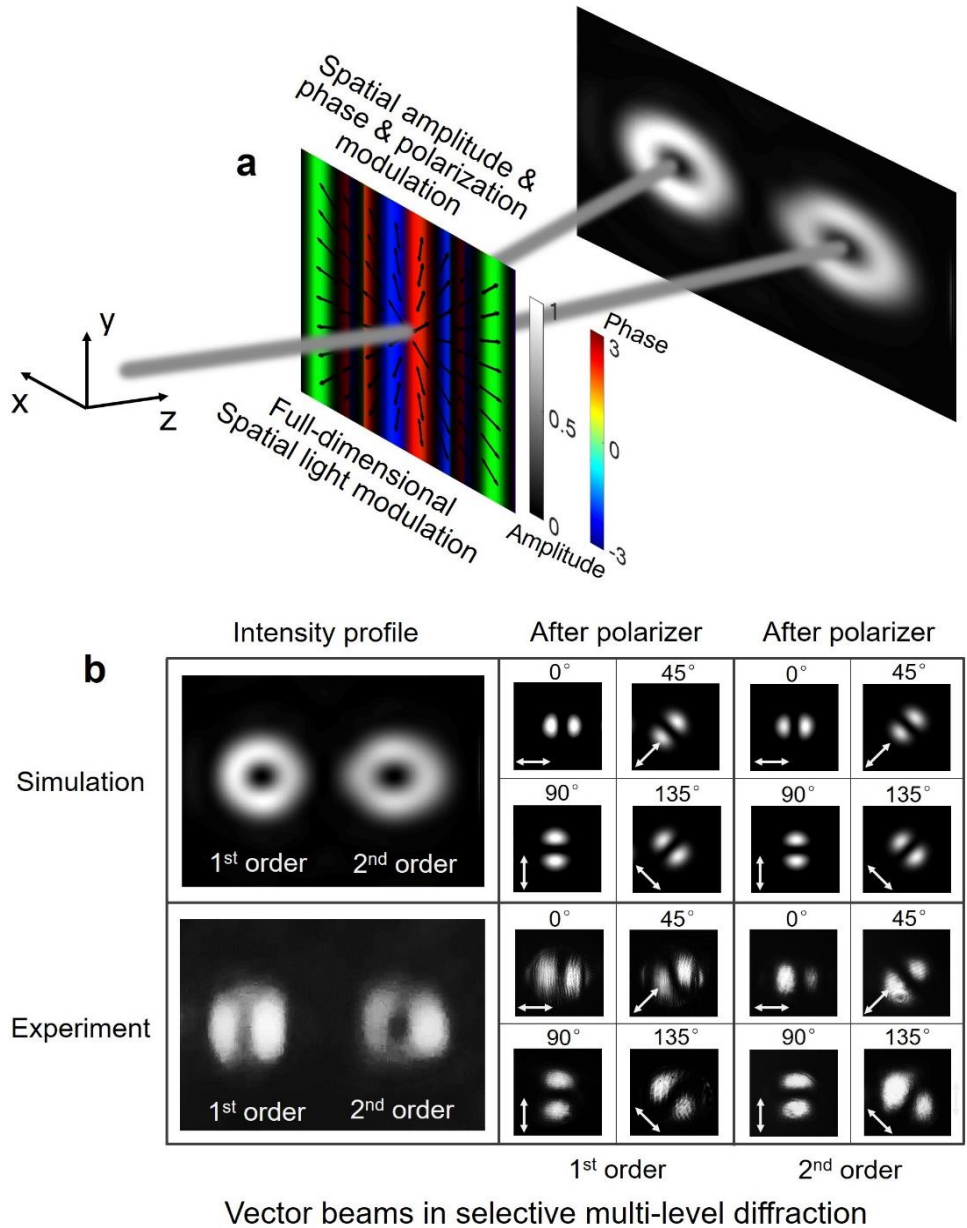

Figure 5. Experimental and simulation results for generating vector beams in selective multi-level diffraction (APB and RPB in the 1st and 2nd diffraction order with equal intensity) with the designed and fabricated full-dimensional spatial light meta-modulator. a Schematic illustration for generating vector beams in selective multi-level diffraction. b Experimental and simulation results for generating vector beams in selective multi-level diffraction (spatial phase, amplitude and polarization dimensions).

## Discussion

Here we discuss some important physics about the full-dimensional spatial light meta-modulators.

First and most important, the central idea of the work, i.e. the parallel tasking, is to divide the total angle of a geometric phase element into multiple sub-angles, each of which can fulfill an independent task. It is emphasized that the full-dimensional spatial light meta-modulator is just one promising application example. In principle the orientation angle of one geometric phase element can be divided into many sub-angles, which may lead to a plethora of opportunities for interesting applications with many different parallel tasks.

Secondly, since the proposed metasurfaces are completely based on the geometric phase element, they are intrinsically broadband in terms of phase modulation. The phase, polarization, and relative amplitude profiles of the produced light beam at the near field of the metasurface transmitted plane are independent of wavelength. By "relative amplitude profile", we remind that the wavelength which can affect the PCE of the geometric phase element will thus homogeneously affect the absolute amplitude of the transmitted beam without changing its relative profile. It means that for the same full-dimensional spatial light meta-modulator, the produced light beams are the same at different wavelengths (only varied with a uniform amplitude factor). However, wavelength is indeed related to the diffraction angle of the produced light beam owing to the structure dispersion, thus different colors can induce different propagation directions. Therefore, chromatic-correction mechanism is needed for multi-color applications such as colored holography and imaging[35–37]. Here, we still describe the metasurface as broadband to emphasize the signature property of the geometric phase in contrast of other type of metasurfaces that modulate light through resonances or effective optical path.

## Conclusion

In summary, we report the parallel-tasking metasurfaces which can modulate the full dimensions of light. The parallel-tasking mechanism employs geometric phase elements while the orientation angle of each element is divided into several sub-angles to introduce different phase modulation. These sub-phases serve as independent control factors (geometric phase, interference, orthogonal superposition) and therefore bestow the designated phase, amplitude and polarization profile to the generated light beam. Under this principle, we design and fabricate plasmonic metasurfaces which produce different types of structured light, and especially the vector beams in a selective multi-level diffraction order as the signature of simultaneous phase, amplitude and polarization modulation. This work demonstrates a flexible, versatile and complete control of light through the parallel-tasking metasurface. The metasurface-enabled full-dimensional spatial light meta-modulators may inspire more interesting applications in optical manipulation, metrology, sensing, imaging, quantum science and optical communications.

## Acknowledgements

This work was supported by the National Key R&D Program of China (2019YFB2203604), the National Natural Science Foundation of China (NSFC) (62125503), the Key R&D Program of Guangdong Province

**Author contributions**

J.W.Z., J.R.Z. and J.W. conceived the idea of the work. J.R.Z. fabricated the devices and carried out the experiments. J.W.Z. and J.W. provided technical supports in device fabrication and experiments. J.W.Z., J.R.Z., Y.J.D and J.W. performed the theoretical analyses, analyzed the data and contributed to writing the paper. J.W.Z. and J.W. finalized the paper. J.W. supervised the project.

**Additional information**

Competing financial interests: The authors declare no competing financial interests.

**Data availability**

All the findings of this study are available in the main text. The raw data are available from the corresponding author upon reasonable request.